# Contextualizing Concepts Using a Mathematical Generalizationof the Quantum Formalism

**Liane M. Gabora and Diederik Aerts**

**ABSTRACT**
We outline the rationale and preliminary results of using the state context property (SCOP) formalism, originally developed as a generalization of quantum mechanics, to describe the contextual manner in which concepts are evoked, used and combined to generate meaning. The quantum formalism was developed to cope with problems arising in the description of (i) the measurement process, and (ii) the generation of new states with new properties when particles become entangled. Similar problems arising with concepts motivated the formal treatment introduced here. Concepts are viewed not as fixed representations, but entities existing in states of potentiality that require interaction with a context—a stimulus or another concept—to 'collapse' to an instantiated form (e.g. exemplar, prototype, or other possibly imaginary instance). The stimulus situation plays the role of the measurement in physics, acting as context that induces a change of the cognitive state from superposition state to collapsed state. The collapsed state is more likely to consist of a conjunction of concepts for associative than analytic thought because more stimulus or concept properties take part in the collapse. We provide two contextual measures of conceptual distance—one using collapse probabilities and the other weighted properties—and show how they can be applied to conjunctions using the pet fish problem.

## 1. INTRODUCTION

Theories of concepts have by and large been *representational theories.* By this we mean that concepts are seen to take the form of fixed mental representations, as opposed to being constructed, or 're-constructed', on the fly through the interaction between the cognitive state and the situation or context.

Representational theories have met with some success. They are adequate for predicting experimental results for many dependent variables including typicality ratings, latency of category decision, exemplar generation frequencies and categorynaming frequencies. However, increasingly, for both theoretical and empirical reasons, they are coming under fire (e.g. Riegler*et al.* 1999, Rosch 1999). As Rosch puts it, they do not account for the fact that concepts have a participatory, not an identifying function in situations. That is, they cannot explain the contextual manner in which concepts are evoked and used (see also Murphy and Medin 1985, Hampton 1987, Medin and Shoben 1988, Gerrig and Murphy 1992, Komatsu 1992). Contextuality is the reason why representational theories cannot describe or predict what happens when two or more concepts arise together, or follow one another, as in the creative generation or interpretation of *conjunctions* of concepts. A concept's meaning shifts depending on what other concepts it arises in the context of (Reed 1972, Storms *et al.* 1996, 1999, Wisniewski 1991, 1997).

This paper shows how formalisms designed to cope with context and conjunction in the microworld may be a source of inspiration for a description of concepts. In this *contextualized*

*theory*,' not only does a concept give meaning to a stimulus or situation, but the situation evokes meaning in the concept, and when more than one is active they evoke meaning in each other.

## 2. Limitations of representational approaches

We begin by briefly summarizing some of the most influential representational theories of concepts, and efforts to delineate what a concept is with the notion of conceptual distance. We then discuss difficulties encountered with representational approaches in predicting membership assessment for conjunctions of concepts. We then show that representational theories have even more trouble coping with the spontaneous emergence or loss of features that can occur when concepts combine.

### 2.1. *Theories of concepts and conceptual distance*

According to the *classical theory* of concepts, there exists for each concept a set of defining features that are singly necessary and jointly sufficient (e.g. Sutcliffe 1993). Extensive evidence has been provided against this theory (or overviews see Smith and Medin 1981, Komatsu 1992).

A number of alternatives have been put forth. According to the *prototype theory* (Rosch 1975, 1978, 1983, Rosch and Mervis 1975), concepts are represented by a set of, not *defining,* but *characteristic* features, which are weighted in the definition of the prototype. A new item is categorized as an instance of the concept if it is sufficiently similar to this prototype. The prototype consists of a set of features $\{a_1, a_2, a_3 ... a_M\}$, with associated weights or applicability values $\{x_1, x_2, x_3 ... x_M\}$, where $M$ is the number of features considered. The distance between a new item and the prototype can be calculated as follows, where $s$ indexes the test stimulus, $x_{sm}$ refers to applicability of $m$th feature to the stimulus $s$, and $x_{pm}$ refers to applicability of $m$th feature to the prototype:

$$d_s = \sqrt{\sum_{m=1}^{M}(x_{sm} - x_{pm})^2} \quad (1)$$

The smaller the value of $d_s$ for a given item, the more representative it is of the concept. Thus concept membership is graded, a matter of degree.

According to the *exemplar theory, (e.g.* Medin *et al.* 1984, Nosofsky 1988, 1992, Heit and Barsalou 1996) a concept is represented by, not defining or characteristic features, but a set of *instances* of it stored in memory. Thus each of the $\{E_1, E_2, E_3, ... E_N\}$ exemplars has a set $\{a_1, a_2, a_3 ... a_M\}$ of features with associated weights $\{x_1, x_2, x_3 ... x_M\}$. A new item is categorized as an instance of concept if it is sufficiently similar to one or more of these previously encountered instances. For example, Storms *et al.* (2000) used the following distance function, where $s$ indexes the test stimulus, $x_{sm}$ refers to applicability of $m$th feature to stimulus $s$, and $x_{nm}$ refers to applicability of $m$th feature to nth most frequently generated exemplar:

$$d_s = \sum_{n=1}^{N}\sqrt{\sum_{m=1}^{M}(x_{sm} - x_{pm})^2} \quad (2)$$

Once again, the smaller the value of ds for a given item, the more representative it is of the concept.

Note that these theories have difficulty accounting for why items that are dissimilar or even opposite might nevertheless belong together; for example, why **white** might be more likely to be categorized with **black** than with **flat,** or why **dwarf** might be more likely to be categorized with **giant** than with say, **salesman.** The only way out is to give the set of relevant 'measurements' or contexts the same status as features, i.e. to lump together as features not only things like 'large' but also things like 'has a size' or 'degree to which size is relevant'.

According to another approach to concepts, referred to as the *theory theory,* concepts take the form of 'mini-theories' (e.g. Murphy and Medin 1985) or schemata (Rummelhart and Norman 1988), in which the causal relationships amongst features or properties are identified. A mini-theory contains knowledge concerning both which variables or measurements are relevant, and the values obtained for them. This does seem to be a step toward a richer understanding of concept representation, though many limitations have been pointed out (see for example Komatsu 1992, Fodor 1994, Rips 1995). Clearly, the calculation of conceptual distance is less straightforward, though to us this reveals not so much a shortcoming of the theory theory, but of the concept of conceptual distance itself. In our view, concepts are not distant from one another at all, but interwoven, and this interwoven structure cannot be observed directly, but only indirectly, as context-specific instantiations. For example, the concept **egg** will be close to **sun** in the context 'sunny side up' but far in the context 'scrambled', and in the context of the DrSuess book *Green Eggs and Ham* it acquires the feature 'green'.

Yet another theory of concepts, which captures their mutable, context-dependent nature, but at the cost of increased vagueness, is *psychological essentialism.* The basic idea is that instances of a concept share a hidden essence which defines its true nature (e.g. Medin and Ortony 1989). In this paper we attempt to get at this notion in a more rigorous and explicit way than has been done.

*2.2. Membership assessments for conjunctive categories*
The limitations of representational theories became increasingly evident through experiments involving conjunctions of concepts. One such anomalous phenomenon is the so-called *guppy effect,* where a guppy is *not* rated as a good example of the concept **pet,** nor of the concept **fish,** but it *is* rated as a good example of **pet fish**(Osherson and Smith 1981).[2] Representational theories cannot account for this. Using the prototype approach, since a guppy is neither a typical pet nor a typical fish, $d_s$ for the guppy stimulus is large for both **pet** and **fish,** which is difficult to reconcile with the empirical result that it is small for **pet fish.** Using the exemplar approach, although a guppy is an exemplar of both **pet** and **fish,** it is unlikely to be amongst the *n* most frequently generated ones. Thus once again $d_s$ is large for both **pet** and **fish,** which is difficult to reconcile with it being small for **pet fish.**

The problem is not solved using techniques from fuzzy set mathematics such as the *minimum rule model,* where the typicality of a conjunction (conjunction typicality) equals the minimum of the typicalities of the two constituent concepts (Zadeh 1965, 1982). (For example, the typicality rating for **pet fish** certainly does not equal the minimum of that for **pet** or **fish.**) Storms *et al.* (2000) showed that a weighted and calibrated version of the minimum rule model can account for a substantial proportion of the variance in typicality ratings for conjunctions exhibiting the guppy effect. They suggested the effect could be due to the existence of *contrast categories,* the idea being that a concept such as **fruit** contains not only information about fruit, but information about categories that are related to, yet different from, fruit. Thus, a particular item might be a better exemplar of the concept **fruit** if it not only has many features in common with exemplars of **fruit** but also few features in common with exemplars of **vegetables** (Rosch and Mervis, 1975). However, another study provided negative evidence for contrast categories (Verbeemen *et al.* in press).

Nor does the theory theory or essence approach get us closer to solving the conjunction problem. As Hampton (1997) points out, it is not clear how a set of syntactic rules for combining or interpreting combinations of mini-theories could be formulated.

## 2.3. *'Emergence' and loss of properties during conjunction*

An even more perplexing problem facing theories of concepts is that, as many studies (e.g. Hastie *et al.* 1990, Kunda *et al.* 1990, Hampton 1997) have shown, a conjunction often possesses features which are said to be *emergent:* not true of its constituents. For example, the properties 'lives in cage' and 'talks' are considered true of **pet birds,** but not true of **pets** or **birds.**

Representational theories are not only incapable of *predicting* what sorts of features will emerge (or disappear) in the conjunctive concept, but they do not even provide a place in the formalism for the gain (or loss) of features. This problem stems back to a limitation of the mathematics underlying not only representational theories of concepts (as well as compositional theories of language) but also classical physical theories. The mathematics of classical physics only allows one to describe a composite or joint entity by means of the product state space of the state spaces of the two subentities. Thus if $X_1$ is the state space of the first subentity, and $X_2$ the state space of the second, the state space of the joint entity is the Cartesian product space $X_1 \times X_2$. For this reason, classical physical theories cannot describe the situation wherein two entities generate a new entity with properties not strictly inherited from its constituents.

One could try to solve the problem *ad hoc* by starting all over again with a new state space each time there appears a state that was not possible given the previous state space; for instance, every time a conjunction like **pet bird** comes into existence. However, this happens every time one generates a sentence that has not been used before, or even uses the same sentence in a slightly different context. Another possibility would be to make the state space infinitely large to begin with. However, since we hold only a small number of items in mind at any one time, this is not a viable solution to the problem of describing what happens in cognition. This problem is hinted at by Boden (1990), who uses the term *impossibilist creativity* to refer to creative acts that no only *explore* the existing state space but *transform* that state space; in other words, it involves the spontaneous generation of new states with new properties.

## 2.4. *The 'obligatory peeking' principle*

In response to difficulties concerning the transformation of concepts, and how mini-theories combine to form conjunctions, Osherson and Smith (1981) suggested that, in addition to a modifiable mini-theory, concepts have a stable definitional *core.* It is the core, they claim, that takes part in the combining process. However, the notion of a core does not straightforwardly solve the conjunction problem. Hampton (1997) suggested that the source of the difficulty is that in situations where new properties emerge during concept conjunction, one is making use of world knowledge, or 'extensional feedback'. He states: 'We cannot expect any model of conceptual combination to account directly for such effects, as they clearly relate to information that is obtained from another source—namely familiarity with the class of objects in the world' (Hampton 1997: 148). Rips (1995) refers to this as the *No Peeking Principle.* Rips' own version of a dual theory distinguishes between representations- of and representations-about, both of which are said to play a role in conjunction. However, he does not claim to have solved the problem of how to describe concepts and their conjunctions, noting 'It seems likely that part of the semantic story will have to include external causal connections that run through the referents and their representations' (Rips 1995: 84).

Goldstone and Rogosky's (in press) ABSURDIST algorithm is a move in this direction. Concept meaning depends on a web of relations to other concepts in the same domain, and the algorithm uses within-domain similarity relations to translate across domains. In our contextualized approach, we take this even further by incorporating not just pre-identified relations amongst concepts, but new relations made apparent in the context of a particular stimulus situation, i.e. the external world. We agree that it may be beyond our reach to predict exactly how world knowledge will come into play in every particular case. However, it is at least possible to put forth a theory of concepts that not only *allows* 'peeking', but in a natural (as opposed to *ad hoc)* way provides a place for it. In fact, in our model, peeking (from either another concept, or an external stimulus) is obligatory; concepts require a peek, a context, to actualize them in some form (even if it is just the most prototypical form). The core or essence of a concept is viewed as a source of potentiality that requires some context to be dynamically actualized, and that thus cannot be described in a context-independent manner (except as a superposition of every possible context-driven instantiation of it). In this view, each of the two concepts in a conjunction constitutes a context for the other that 'slices through' it at a particular angle, thereby mutually actualizing one another's potentiality in a specific way. As a metaphorical explanatory aid, if concepts were apples, and the stimulus a knife, then the qualities of the knife would determine not just which apple to slice, but which direction to slice through it. Changing the knife (the context) would expose a different face of the apple (elicit a different version of the concept).And if the knife were to slash through several apples (concepts) at once, we might end up with a new kind of apple (a conjunction).

## 3.Two cognitive modes: analytic and associative
We have seen that, despite considerable success when limited to simple concepts like **bird,** representational theories run into trouble when it comes to conjunctions like **pet bird** or even **green bird.** In this section we address the question: why would they be so good for modeling many aspects of cognition, yet so poor for others?

### 3.1. *Creativity and flat associative hierarchies*
It is widely suggested that there exist two forms of thought (e.g. James 1890, Piaget 1926, Neisser 1963, Johnson-Laird 1983, Dennett 1987, Dartnell 1993, Sloman 1996, Rips 2001a). One is a focused, evaluative *analytic mode,* conducive to analysing relationships of *cause and effect.* The other is an intuitive creative *associative mode* that provides access to remote or subtle connections between features that may be *correlated* but not necessarily causally related. We suggest that while representational theories are fairly adequate for predicting and describing the results of cognitive processes that occur in the analytical mode, their shortcomings are revealed when it comes to predicting and describing the results of cognitive processes that occur in the associative mode, due to the more contextual nature of cognitive processes in this mode.

Since the associative model is thought to be more evident in creative individuals, it is useful at this point to look briefly at some of the psychological attributes associated with creativity. Martindale (1999) has identified a cluster of such attributes, including defocused attention (Dewing and Battye 1971, Dykes and McGhie 1976, Mendelsohn 1976), and high sensitivity (Martindale and Armstrong 1974, Martindale 1977), including sensitivity to subliminal impressions, that is, stimuli that are perceived but of which we are not *conscious* of having perceived (Smith and Van de Meer 1994).

Another characteristic of creative individuals is that they have *flat associative hierarchies* (Mednick 1962). The steepness of an individual's associative hierarchy is measured experimentally by comparing the number of words that individual generates in response to stimulus words on a word association test. Those who generate only a few words in response to the stimulus have a *steep* associative hierarchy, whereas those who generate many have a *flat* associative hierarchy. Thus, once such an individual has run out of the more usual associations (e.g. **chair** in response to **table),** unusual ones (e.g. elbow in response to **table)** come to mind.

It seems reasonable that in a state of defocused attention and heightened sensitivity, more features of the stimulus situation or concept under consideration get processed. (In other words, the greater the value of *M* in equations (1) and (2) for prototype and exemplar theories.) It also seems reasonable that flat associative hierarchies result from memories and concepts being more richly etched into memory; thus there is a greater likelihood of an associative link between any two concepts. The experimental evidence that flat associative hierarchies are associated with defocused attention and heightened sensitivity suggests that the more features processed, the greater the potential for associations amongst stored memories and concepts. We can refer to the detail with which items are stored in memory as *associative richness*.

**3.2.** *Activation of conceptual space: spiky versus flat*
We now ask: how might different individuals, or a single individual under different circumstances, vary with respect to degree of detail with which the stimulus or object of thought gets etched into memory, and resultant degree of associative richness?[3] Each memory location is sensitive to a broad range of features, or values of an individual feature (e.g., Churchland and Sejnowski 1992). Thus although a particular location responds maximally to lines of a certain orientation, it may respond somewhat to lines of a close orientation. This is referred to as *coarse coding*. It has been suggested that the coarseness of the coding—that is, the size of the higher cortical receptive field—changes in response to attention (Kruschke 1993). Kruschke's neural network model of categorization, ALCOVE, incorporates a selective attention mechanism, which enables it to vary the number of dimensions the network takes into account at a time, and thereby mimics some previously puzzling aspects of human categorization. In neural networks, receptive field activation can be graded using a radial basis function (RBF). Each input activates a hypersphere of hidden nodes, with activation tapering off in all directions according to a (usually) Gaussian distribution of width $\sigma$(Willshaw and Dayan, 1990, Hancock *et al.,* 1991, Holden and Niranjan, 1997, Lu *et al.* 1997).[4] Thus if a is small, the input activates a few memory locations but these few are hit hard; we say the activation function is *spiky*. If $\sigma$ is large, the input activates many memory locations to an almost equal degree; we say the activation function is relatively *flat*.

Whether or not human memory works like a RBF neural network, the idea underlying them suggests a basis for the distinction between associative and analytic modes of thought. We will use the terms spiky and flat activation function to refer to the extent to which memory gets activated by the stimuli or concepts present in a given cognitive state, bearing in mind that this may work differently in human cognition than in a neural network.[5] The basic idea then is that when the activation function is spiky, only the most typical, central features of a stimulus or concept are processed. This is conducive to analytic thought where remote associations would be merely a distraction; one does not want to get sidetracked by features that are atypical, or modal (Rips 2001b), which appear only in imagined or counterfactual instances. However, as the number of features or stimulus dimensions increases, features that are less central to the concept that best categorizes it start to get included, and these features may in fact make it defy straightforward classification as strictly an instance of one concept or another. When the activation function is relatively flat, more features are attended and participate in the process of activating and evoking from memory; atypical as well as typical ones. Therefore, more memory locations participate in the release of 'ingredients' for the next instant. These locations will have previously been modified by (and can therefore be said to 'store' in a distributed manner) not only concepts that obviously share properties with the stimulus, but also concepts that are correlated with it in unexpected ways. A flat activation function is conducive to creative, associative thought because it provides a high probability of evoking one or more concepts not usually associated with the stimulus.

Thus we propose that representational theories—in which concepts are depicted as fixed sets of attributes—are adequate for modeling analytical processes, which establish relationships of cause and effect amongst concepts in their most prototypical forms. However, they are not adequate for modeling associative processes, which involve the identification of correlations amongst more richly detailed, context-specific forms of concepts. In the associative mode, aspects of a situation the relevance of which may not be readily apparent, or relations to other concepts which have gone unnoticed—perhaps of an analogical or metaphorical nature—can 'peek through'. A cognitive state in which a new relationship amongst concepts is identified is a state of *potentiality,* in the sense that the newly identified relationship could be resolved different ways depending on the contexts one encounters, both immediately, and down the road. For example, consider the cognitive state of the person who thought up the idea of building a snowman. It seems reasonable that this involved thinking of snow not just in terms of its most typical features such as 'cold' and 'white', but also the less typical feature 'moldable'. At the instant of inventing **snowman** there were many ways of resolving how to give it a nose. However, perhaps because the inventor happened to have a carrot handy, the concept **snowman** has come to acquire the feature 'carrot nose'.

## 4. A formalism that incorporates context

We have seen that models of cognition have difficulty describing contextual, associative, or correlation-based processes. This story has a precedent. Classical physics does exceedingly well at describing and predicting relationships of *causation,* but it is much less powerful in dealing with results of experiments that entail sophisticated relationships of *correlation.* It cannot describe certain types of correlations that appear when quantum entities interact and combine to form joint entities. According to the dynamical evolution described by the Schrodinger equation, whenever there is interaction between quantum entities, they spontaneously enter an entangled state that contains new properties that the original entities did not have. The description of this birth of new states and new properties required the quantum mechanical formalism.

Another way in which the shortcomings of classical mechanics were revealed had to do in a certain sense with the issue of 'peeking'. A quantum particle could not be observed without disturbing it; that is, without changing its state. Classical mechanics could describe situations where the effect of a measurement was negligible, but not situations where the measurement intrinsically influenced the evolution of the entity. The best it could do is avoid as much as possible any influence of the measurement on the physical entity under study. As a consequence, it had to limit its set of valuable experiments to those that have almost no effect on the physical entity (called observations). It could not incorporate the context generated by a measurement directly into the formal description of the physical entity. This too required the quantum formalism.

In this section we first describe the pure quantum formalism. Then we briefly describe the generalization of it that we apply to the description of concepts.

### 4.1. *Pure quantum formalism*
In quantum mechanics, the state of a physical entity can change in two ways: (i) under the influence of a measurement context, and this type of change is called *collapse,* and (ii) under the influence of the environment as a whole, and this change is called *evolution.* A state $\psi$ is represented by a unit vector of a complex Hilbert space $\mathcal{H}$, which is a vector space over the complex numbers equipped with an inproduct (see Appendix I). A property of the quantum entity is described by a closed subspace of the complex Hilbert space or by the orthogonal projection operator $P$ corresponding to this closed subspace, and a measurement context by a self-adjoint operator on the Hilbert space, or by the set of orthogonal projection operators that constitute the spectral family of this self-adjoint operator (see Appendix II). If a quantum entity is in a state of $\psi$, and a measurement context is applied to it, the state $\psi$ changes to the state:

$$\frac{P(\psi)}{\|P(\psi)\|} \tag{3}$$

where $P$ is the projector of the spectral family of the self-adjoint operator corresponding to the outcome of the measurement. This change of state is more specifically what is meant by the term collapse. It is a probabilistic change and the probability for state $\psi$ to change to state $(P(\psi))/\|P(\psi)\|$ under the influence of the measurement context is given by:

$$\langle \psi, P(\psi) \rangle \tag{4}$$

where $\langle , \rangle$ is the inproduct of the Hilbert space (see Appendix II).

The state prior to, and independent of, the measurement, can be retrieved as a theoretical object—the unit vector of complex Hilbert space that reacts to all possible measurement contexts in correspondence with experimental results. One of the merits of quantum mechanics is that it made it possible to describe the undisturbed and unaffected state of an entity even if most of the experiments needed to measure properties of this entity disturb this state profoundly (and often even destroy it). In other words, the message of quantum mechanics is that it is possible to describe a reality that only can be known through acts that alter this reality.

There is a distinction in quantum mechanics between similarity in terms of which measurements or contexts are relevant, and similarity in terms of values for these measurements (a distinction which we saw in section two has not been present in theories of concepts). Properties for which the same measurement—such as the measurement of spin—is relevant are said to be *compatible* with respect to this measurement. One of the axioms of quantum mechanics—called *weak modularity*— is the requirement that orthogonal properties—such as 'spin up' and 'spin down'— are compatible.

In quantum mechanics, the conjunction problem is seriously addressed, and to some extent solved, as follows. When quantum entities combine, they do not stay separate as classical physical entities tend to do, but enter a state of *entanglement.* If $\mathcal{H}_1$ is the Hilbert space describing a first subentity, and $\mathcal{H}_2$ the Hilbert space describing a second subentity, then the joint entity is described in the tensor product space $\mathcal{H}_1 \otimes \mathcal{H}_2$ of the two Hilbert spaces $\mathcal{H}_1$ and $\mathcal{H}_2$. The tensor product always allows for the emergence of new states—specifically the entangled states—with new properties.

The presence of entanglement—*i.e.* quantum structure—can be tested for by determining whether correlation experiments on the joint entity violate Bell inequalities (Bell 1964). Pitowsky (1989) proved that if Bell inequalities are satisfied for a set of probabilities concerning the outcomes of the considered experiments, there exists a classical Kolmogorovian probability model that describes these probabilities. The probability can then be explained as being due to a lack of knowledge about the precise state of the system. If, however, Bell inequalities are violated, Pitowsky proved that no such classical Kolmogorovian probability model exists. Hence, the violation of Bell inequalities shows that the probabilities involved are non-classical. The only type of non-classical probabilities that are well known in nature are the quantum probabilities.

**4.2.*Generalized quantum formalism***
The standard quantum formalism has been generalized, making it possible to describe changes of state of entities with any degree of contextuality, whose structure is not purely classical nor purely quantum, but something in between (Mackey 1963, Jauch 1968, Piron 1976, 1989, 1990, Randall and Foulis 1976, 1978, Foulis and Randall 1981, Foulis *et al.* 1983, Pitowsky 1989, Aerts 1993, 2002, Aerts and Durt 1994 a, b). The generalizations of the standard quantum formalism have been used as the core mathematical structure replacing the Hilbert space of standard quantum mechanics the structure of *a lattice*, representing the set of features or properties of the physical entity under consideration. Many different types of lattices have been introduced, depending on the type of generalized approach and on the particular problem under study. This has resulted in mathematical structures that are more elaborate than the original lattice structure, and it is one of them, namely the *state context property system*, or *SCOP*, that we take as a starting point here.

Let us now outline the basic mathematical structure of a *SCOP*. It consists of three sets and two functions, denoted:

$$(\Sigma, \mathcal{M}, \mathcal{L}, \mu, \nu) \tag{5}$$

where: $\Sigma$ is the set of possible states; $\mathcal{M}$ is the set of relevant contexts; $\mathcal{L}$ is the lattice which describes the relational structure of the set of relevant properties or features; $\mu$ is a probability function that describes how a couple $(e, p)$, where $p$ is a state, and $e$ a context, transforms to a couple $(f, q)$ where $q$ is the new state (collapsed state for context $e$), and $f$ the new context; $\nu$ is the weight or applicability of a certain property, given a specific state and context. The structure $\mathcal{L}$ is that of a complete, orthocomplemented lattice. This means that:

- A partial order relation denoted $<$ on $\mathcal{L}$ representing that the implication of properties, i.e. actualization of one property implies the actualization of another. For $a, b \notin \mathcal{L}$ we have:

$$a < b \Leftrightarrow \text{if } a \text{ then } b \tag{6}$$

- Completeness: infimum (representing the conjunction and denoted ∧) and supremum (representing the disjunction and denoted ∨) exists for any subset of properties. 0, minimum element, is the infimum of all elements of $\mathcal{L}$ and $I$, maximal element, is the supremum of all elements of $\mathcal{L}$.

- Orthocomplemented: an operation⊥ exists, such that for $a, b \notin \mathcal{L}$ we have:

$$(a^\perp)^\perp = a \qquad (7)$$

$$a < b \Rightarrow b^\perp < a^\perp \qquad (8)$$

$$a \wedge a^\perp = 0, \quad a \vee a^\perp = 1 \qquad (9)$$

Thus $a^\perp$ is the 'negation' of $a$.

- Elements of $\mathcal{L}$ are weighted. Thus for state, $p$, context $e$ and property $a$ there exists weight $v(p, e, a)$, and for a $a \notin \mathcal{L}$:

$$v(p, e, a) + v(p, e, a^\perp) = 1 \qquad (10)$$

These general formalisms describe much more than is needed for quantum mechanics, and in fact, standard quantum mechanics and classical mechanics fall out as special cases (Aerts 1983). For the *SCOP* description of a pure quantum entity, see Appendix III.

    It is gradually being realized that the generalized quantum formalisms have relevance to the macroscopic world (e.g. Aerts 1991, Aerts, Aertset a/. 2000, Aerts, Broekaert*et al.* 2000). Their application beyond the domain that originally gave birth to them is not as strange as it may seem. It can even be viewed as an unavoidable sort of evolution, analogous to what has been observed for chaos and complexity theory. Although chaos and complexity theory were developed for application in inorganic physical systems, they quickly found applications in the social and life sciences, and are now thought of as domain-general mathematical tools with broad applicability. The same is potentially true of the mathematics underlying the generalized quantum formalisms. Although originally developed to describe the behavior of entities in the microworld, there is no reason why their application should be limited to this realm. In fact, given the presence of potentiality and contextuality in cognition, it seems natural to look to these formalisms for guidance in the development of a formal description of cognitive dynamics.

## 5. Application of SCOP to concepts

In this section we apply the generalized quantum formalism—specifically the SCOP—to cognition, and show what concepts reveal themselves to be within this framework. To do this we must make a number of subtle but essential points. Each of these points may appear strange and not completely motivated in itself, but together they deliver a clear and consistent picture of what concepts are.

    We begin by outlining some previous work in this direction. Next we present the mathematical framework. Then we examine more closely the roles of potentiality, context, collapse and actualization. Finally we will focus more specifically on how the formalism is used to give a measure of conceptual distance. This is followed up in the next section, which shows

using a specific example how the formalism is applied to concept conjunction.

### 5.1. *Previous work*
One of the first applications of these generalized formalisms to cognition was modeling the decision making process. Aerts and Aerts (1994) proved that in situations where one moves from a state of indecision to a decided state (or vice versa), and the change of state is context-dependent, the probability distribution necessary to describe it is non-Kolmogorovian. Therefore a classical probability model cannot be used. Moreover, they proved that such situations *can* be accurately described using these generalized quantum mathematical formalisms. Their mathematical treatment also applies to the situation where a cognitive state changes in a context-dependent way to an increasingly specified conceptualization of a stimulus or situation. Once again, context induces a non-deterministic change of the cognitive state that introduces a non-Kolmogorivian probability on the state space. Thus, a non-classical (quantum or generalized quantum) formalism is necessary.

Using an example involving the concept cat and instances of cats, we proved that Bell inequalities are violated in the relationship between a concept and specific instances of it (Aerts, Aerts *et al*. 2000). Thus we have evidence that this formalism reflects the underlying structure of concepts. In (Aerts, D'Hondt *et al.* 2000) we show that this result is obtained because of the presence of EPR-type correlations amongst the features or properties of concepts. The EPR nature of these correlations arises because of how concepts exist in states of potentiality, with the presence or absence of particular properties being determined *in the process* of the evoking or actualizing of the concept. In such situations, the mind handles disjunction in a quantum manner. It is to be expected that such correlations exist not only amongst different instances of a single concept, but amongst different related concepts, which makes the notion of conceptual distance even more suspect.

### 5.2. *Mathematical framework*
In the development of this approach, it became clear that to be able to describe contextual interactions and conjunctions of concepts, it is useful to think not just in terms of concepts *per se*, but in terms of the cognitive states that instantiate them. Each concept is potentially instantiated by many cognitive states; in other words, many thoughts or experiences are interpreted in terms of any given concept. This is why we first present the mathematical structure that describes an entire conceptual system, or mind. We will then illustrate how concepts appear in this structure. We use the mathematical structure of a state context property system or SCOP:

$$(\Sigma, \mathcal{M}, \mathcal{L}, \mu, \nu) \qquad (11)$$

where: $\Sigma$ is the set of all possible cognitive states, sometimes referred to as conceptual space. We use symbols $p, q, r, ...$ to denote states; $\mathcal{M}$ is the set of relevant contexts that can influence how a cognitive state is categorized or conceptualized. We use symbols $e, f, g, ...$ to denote contexts; $\mathcal{L}$ is the lattice which describes the relational structure of the set of relevant properties or features. We use symbols $a, b, c, ...$ to denote features or properties; $\mu$ is a property function that describes how a couple $(e, p)$—where $p$ is a state, and $e$ a context—transforms to a couple $(f, q)$, where $q$ is the new state (collapsed state for context $e$), and $f$ the new context; $\nu$ is the weight or applicability of a certain property, given a specific state and context.

By cognitive states we man states of the mind (the mind being the entity that experiences them). Whereas the two sets $\Sigma$ and $\mathcal{M}$, along with the function $\mu$, constitute the set of possible cognitive states and the contexts that evoke them, the set $\mathcal{L}$ and the function $\nu$, describe properties of these states, and their weights. In general, a cognitive state $p \in \Sigma$ under context $e$ (the stimulus) changes to state $q \in \Sigma$ according to probability function $\mu$. Even if the stimulus situation itself does not change, the change of state from $p$ to $q$ changes the context (i.e. the stimulus is now experienced in the context of having influenced the change of state from $p$ and $q$). Thus we have a new context $f$. For a more detailed exposition of SCOP applied to cognition, see appendix D.

### 5.3. *How concepts appear in the formalism*
We denote concepts by the symbols $A, B, C,...$ and the set of all concepts $\mathcal{A}$. A concept appears in the formalism as a subentity of this entire cognitive system, the mind.[6] This means that if we consider a cognitive state $p \in \Sigma$, for each concept $A \in \mathcal{A}$, there exists a corresponding state $p_A$ of this concept. The concept $A \in \mathcal{A}$ is described by its own SCOP denoted $(\Sigma_A, \mathcal{M}, \mu_A, \mathcal{L}_A, \nu_A)$, where $\Sigma_A$ is the set of states of this concept, and $\mathcal{M}$ is the set of contexts. Remark that $\mathcal{M}$ is the same for different concepts, and for the mind as a whole, because all contexts that are relevant for the mind as a whole are also relevant for a single concept. Furthermore, $\mu_A$ describes the probabilities of collapse between states and contexts for this concept, and $\mathcal{L}_A$ and $\nu_A$ refer to the set of features and weights relevant to concept $A$. When we speak of the potentiality of a concept, we refer to the totality of ways in which it could be actualized, articulated, or experienced in a cognitive state, given all the different contexts in which it could be relevant.

5.3.1. *Instantiation of concept actualizes potential.* For a set of concepts $\{A_1, A_2,...,A_n,...\}$, where $A_i \in \mathcal{A} \forall i$, the cognitive state $p$ can be written $\{p_{A_1}, p_{A_2}, p_{A_3},..., p_{A_n},...\}$, where each $p_{A_i}$ is a state of concept $A_i$. For a given context $e \in \mathcal{M}$, each of these states $p_A$ could be a potentiality state or a collapsed state. Let us consider the specific situation where the cognitive state $p$ instantiates concept $A_m$. What this explicitly means is that $p_{A_m}$, the state of concept $A_m$ becomes an actualized cognitive state, and this corresponds to the evoking of concept $A_m$. At the instant $A_m$ is evoked in cognitive state $p$, its potentiality is momentarily deflated or collapsed with respect to the given context $e$.

5.3.2. *Uninstantiated concepts retain potential.* Let us continue considering the specific situation where state p instantiates concept $A_m$ under context $e$. For each concept $A_i$ where $i \neq m$, no instantiation takes place, and state $p_{A_i}$ remains a complete potentially state for context $e$. Thus, concepts that are not evoked in the interpretation of a stimulus to become present in the cognitive state retain their potentiality. This means they are not limited to a fixed set of features or relations amongst features. The formalism allows for this because the state space where a concept 'lives' is not limited a *priori* to features thought to be relevant. It is this that allows both their contextual character to be expressed, with new features emerging under new contexts. Given the right context were to come along, any feature could potentially become incorporated into an instantiation of it.

5.3.3. *Concepts as contexts and features.* In addition to appearing as subentities instantiated by cognitive states, concepts appear in the formalism in two other ways. First, they can constitute (part of) a context $e \in \mathcal{M}$. Second, something that constitutes a feature or property $a \in \mathcal{L}$ in one situation can constitute a concept in another; for instance, 'blue' is a property of the sky, but also one has a concept **blue.** Thus, the three sets $\Sigma$, $\mathcal{M}$ and $\mathcal{L}$, of a SCOP are all in some way affected by concepts.

5.3.4. *Conjunctions of concepts.* As mentioned previously, the operation applied to pure quantum entities is the tensor product. The algebraic operation we feel to be most promising for the description of conjunction of concepts is the following. In a SCOP, there is a straightforward connection between the state of the entity under consideration at a certain moment, and the set of properties that are actual at that moment.[7] This makes it possible to, for a certain fixed property $a \in \mathcal{L}$, introduce what is called the relative SCOP for $a$, denoted $(\Sigma, \mathcal{M}, \mathcal{L}, \mu, \nu)_a$. Suppose that $(\Sigma, \mathcal{M}, \mathcal{L}, \mu, \nu)$ describes concept $A$, then $(\Sigma, \mathcal{M}, \mathcal{L}, \mu, \nu)_a$ describes concept $A$ given that property $a$ is always actual for $A$. We could, for example, describe with this structure the concept **pet** where the property *swims* is always actual. This would give us a possible model for the conjunction of a noun concept with an adjective concept. In the case of **pet** and *swims* this would come close to **pet fish,** but of course, that this happens is certainly not a general rule. For the case of a conjunction of two nouns, if we want to try out the relative SCOP construction, we would have to consider the conjunctions of all possible features of the two nouns and derive from this the SCOP that would describe the conjunction of the two nouns.

## 5.4. *Superposition, potentiality couples and change of cognitive state*
We cannot specify with complete accuracy (i) the content of state *p*, nor (ii) the stimulus situation it faces, context *e*, nor (iii) how the two will interact. Therefore, any attempt to mathematically model the transition from *p* to *q* must incorporate the possibility that the situation could be interpreted in many different ways, and thus many different concepts (or conjunctions of them) being activated. Within the formalism, it is the structure of the probability field $\mu$ that describes this. For a given state *p*, and another state $q \in \Sigma$ and contexts *e* and $f \in \mathcal{M}$, the probability $\mu(f, q, e, p)$ that state *p* change *s* under the influence of context *e* to state *q* (and that *e* changes to *f*) will often be different from zero. In the quantum language, we can express this by saying that p is a superposition state of all the states $q \in \Sigma$ such that the probability $\mu(f, q, e, p)$ is non-zero for some $e, f \in \mathcal{M}$. Note that whether or not *p* is in a state of potentiality depends on the context *e*. It is possible that state *p* would be a superposition state for *e* but not for another context *f*. Therefore, we use the term *potentiality couple (e, p)*.

We stress that the potentiality couple is different from the potentiality of a concept; the potentiality couple refers to the cognitive state (in all its rich detail) with respect to a certain context (also in all its rich detail), wherein a particular instantiation of some concept (or conjunction of them) may be what is being subjectively experienced. However, they are related in the sense that the potentiality of *p* decreases if concepts $A \in \mathcal{A}$ evoked in it enter collapsed states.

5.4.1. *Collapse: non-deterministic change of cognitive state.* Following the quantum terminology, we refer to the cognitive state following the change of state under the influence of a context as a collapsed state. Very often, though certainly not always, a state *p* is a superposition state with respect to context *e* and it collapses to state *q* which is an eigenstate[8] with respect to *e*, but a superposition state with respect to the new context *f*. This is the case when couple (*e*,*p*) refers to conception of stimulus prior to categorization, and couple (*f*, *q*) refers to the new situation after categorization has taken place.

Recall that a quantum particle cannot be observed or 'peeked at' without disturbing it;

that is, without inducing a change of state. Similarly, we view concepts as existing in states of potentiality which require a context—activation by a stimulus or other concept that constitutes (perhaps partially) the present cognitive state—to be elicited and thereby constitute the content (perhaps partially) of the next cognitive state. However, just as in the quantum case, this 'peeking' causes the concept to collapse from a state of potentiality to a particular context-driven instantiation of it. Thus, the stimulus situation plays the role of the measurement by determining which are the possible states that can be collapsed upon; it 'tests' in some way the potentiality of the associative network, forces it to actualize, in small part, what it is capable of. A stimulus is categorized as an instance of a specific concept according to the extent to which the conceptualization or categorization of it constitutes a context that collapses the cognitive state to a thought or experience of the concept.

5.4.2. *Deterministic change of cognitive state.* A special case is when the couple (*e, p*) is *not* a potentiality couple. This means there exists a context *f* and a state *q*, such that with certainty couple (*e, p*) changes to couple (*f, q*). In this case we call (*e, p*) a deterministic couple and *p* a deterministic state as a member of the couple (*e, p*). An even more special case is when the context *e* does not provoke any change of the state *p*. Then the couple (*e, p*) is referred to as an eigencouple, and the state *p* an eigenstate as a member of the couple (*e, p*).

5.4.3. *Retention of potentiality during collapse.* For a given stimulus *e*, the probability that the cognitive state *p* will collapse to a given concept *A* is related to the algebraic structure of the total state context property system ($\Sigma, \mathcal{M}, \mathcal{L}, \mu, \nu$) and most of all, to the probability field $\mu$(*f, q, e, p*) that describes how the stimulus and the cognitive state interact. It is clear that, much as the potentiality of a concept (to be applicable in all sorts of contexts) is reduced to a single actualized alternative when it collapses to a specific instantiation, the potentiality of a stimulus (to be interpreted in all sorts of ways) is diminished when it is interpreted in terms of a particular concept. Thus, in the collapse process, the stimulus loses potentiality. Consider as an example the situation that one sees a flower, but if one were to examine it more closely, one would see that it is a plastic flower. One possibility for how a situation such as this gets categorized or conceptualized is that extraneous, or modal, feature(s) are discarded, and the cognitive state collapses completely to the concept that at first glance appears to best describe it: in this case, **flower.** We can denote this cognitive state $p_1 \in \Sigma$ Some of the richness of the particular situation is discarded, but what is gained is a straightforward way of framing it in terms of what has come before, thus immediately providing a way to respond to it: as one has responded to similar situations in the past. This is more likely if one is in an analytical mode and thus $\sigma$ is small, such that one does not encode subtle details (e.g. 'the flower is made of plastic').

However, a stimulus may be encoded in richer detail such that, in addition to features known to be associated with the concept that could perhaps best describe it, atypical or modal features are encoded. This is more likely if one is in an associative mode, and thus $\sigma$ is large. Let us denote as $P_2 \in \Sigma$ the state of perceiving something that is flower-like, but that appears to be 'made of plastic'. The additional feature(s) of $P_2$ may make it more resistant to immediate classification, thereby giving it potentiality. In the context of wanting to make a room more cheerful it may serve the purpose of a flower, and be treated as a flower, whereas in the context of a botany class it will not. The state $P_2$, that retains potentiality may be close to $P_l$, the completely collapsed state, but not identical to it. In general, the flatter the activation function, the more features of the stimulus situation are perceived and thus reflected to and back from the associative network. Thus the more likely that some aspects of the situation do not fall cleanly into any particular category or concept, and therefore the more potentially present in the cognitive state, and the more non-classical the reflection process. Note that in an associative mode, for a given cognitive state there will be more features to be resolved, and so the variety of potential ways of collapsing will tend to be greater. Hence the set of states that can be collapsed to is larger.

5.4.4. *Loss of potentiality through repeated collapse.* It seems reasonable that the presence of potentiality in a cognitive state for a certain context is what induces the individual to continue thinking about, recategorizing, and reflecting on the stimulus situation. Hence if the cognitive state is like $p_2$, and some of the potentiality of the previous cognitive state was retained, this retained potentiality can be collapsed in further rounds of reflecting. Thus a stream of collapse ensues, and continues until the stimulus or situation can be described in terms of, not just one concept (such as **flower),** complex conjunction of concepts (such as 'this flower is made of plastic so it is not really a flower'). This is a third state $p_3$, that again is a collapsed state, but of a more complex nature than the first collapsed state $p_1$ was. But it is more stable with respect to the stimulus than $p_1$ or $p_2$.

The process can continue, leading to a sequence of states $P_3, P_4, P_5,...$. With each iteration the cognitive state changes slightly, such that over time it may become possible to fully interpret the stimulus situation in terms of it. Thus, the situation eventually gets interpreted as an instance of a new, more complex concept or category, formed spontaneously through the conjunction of previous concepts or categories during the process of reflection. The process is contextual in that it is open to influence by those features that did not fit the initial categorization, and by new stimuli that happen to come along.

5.5. *Contextual conceptual distance*

We have claimed that for any concept, given the right context, any feature could potentially become involved in its collapse, and thus the notion of conceptual distance becomes less meaningful. However, it is possible to obtain a measure of the distance between states of concepts, potentiality states as well as collapsed states (which can be prototypes, exemplars, or imaginary constructions), and this is what the formulas here measure.

5.5.1. *Probability conceptual distance.* First, we define what we believe to be the most direct distance measure, based on the probability field $\mu(f, q, e, p)$. This method is analogous to the procedure used for calculating distance in quantum mechanics. We first introduce a reduced probability:

$$\mu: \Sigma \times \mathcal{M} \times \Sigma \to [0,1] \tag{12}$$

$$(q, e, p) \mapsto \mu(q, e, p) \tag{13}$$

where:

$$\mu(q, e, p) = \sum_{f \in \mathcal{M}} \mu(f, q, e, p) \tag{14}$$

and $\mu(q, e, p)$ is the probability that state $p$ changes to state $q$ under the influence of context $e$.

The calculation of probability conceptual distance is obtained using a generalization of the distance in complex Hilbert space for the case of a pure quantum situation, as follows:

$$d_\mu(q, e, p) = \sqrt{2(1 - \sqrt{\mu(q,e,p)})} \tag{15}$$

We can also introduce the conceptual angle between two states, again making use of the formula from pure quantum mechanics:

$$\theta_\mu(q, e, p) = \arccos \mu(q, e, p) \tag{16}$$

We call $d_\mu$ the probability conceptual distance, or the $\mu$ distance, and $\theta_\mu$ the probability conceptual angle, or the $\mu$ angle. For details, see appendix A and equations (31) and (32), and remark that for unit vectors (31) reduces to (15).

Let us consider some special cases to see more clearly what is meant by this distance and this angle. If $\mu(q, e, p) = 0$ we have $d_\mu(q, e, p) = \sqrt{2}$ and $\theta_\mu(q, e, p) = \frac{\pi}{2}$. This corresponds to the distance and angle between two orthogonal unit vectors in a vectorspace. So orthogonality of states, when the probability that one state changes to the other state is 0, represent the situation where the distance is maximal ($\sqrt{2}$), and the angle is a straight angle ($\frac{\pi}{2}$). If $\mu(q, e, p) = 1$, we have $d_\mu(q, e, p) = 0$ and $\theta_\mu(q, e, p) = 0$. This corresponds to the distance and angle between two coinciding unit vectors in a vectorspace. So coincidence of states—when the probability that one state changes to the other state = 1—represents the situation where the distance is minimal (0), and the angle is minimal (0). For values of $\mu(q, e, p)$ strictly between 0 and 1, we find a distance between 0 and $\sqrt{2}$ and an angle between 0 and $\frac{\pi}{2}$.

It is important to remark that the distance $d_\mu(q, e, p)$ and angle $\theta_\mu(q, e, p)$ between two states $p$ and $q$ is dependent on the context $e$ that provokes the transition from $p$ to $q$. Even for a fixed context, the distance does not necessarily satisfy the requirements that a distance is usually required to satisfy in mathematics. For example, it is not always the case that $d_\mu(q, e, p) = (d_\mu(p, e, q))$, because the probability $\mu(q, e, p)$ for $p$ to change to $q$ under context $e$ is not necessarily equal to the probability $\mu(p, e, q)$ for $q$ to change to $p$ under context $e$.[9]

5.5.2. *Property conceptual distance.* In order to illustrate explicitly the relationship between our approach and the distance measures provided by the prototype and exemplar approaches described previously, we define a second distance measure based on properties. This distance measure requires data on the probability of collapse of a cognitive state under the influence of a context to a cognitive state in which a particular feature is activated. In order to define operationally what this data refers to, we describe how it could be obtained experimentally. One group of subjects is asked to consider one particular concept $A$, and this evokes in them cognitive state $p$. This state will be subtly different for each subject, depending on the specific contexts which led them to form these concepts, but there will be nevertheless commonalities. A second group of subjects is asked to consider another concept $B$, which evokes cognitive state $q$. Again, $q$ will be in some ways similar and in some ways different for each of these subjects. The subjects are then asked to give an example of 'one' feature for each one of the considered concepts. Thus, two contexts are at play: context $e$ that consists of asking the subject to give a feature of the concept focused on in state $p$, and context $f$ that consists of asking the subject to give a feature of the concept focused on in state $q$. Thus we have two potentiality couples $(e, p)$ and $(f, q)$. Suppose couple $(e,p)$ gives rise to the list of features $\{b_1, b_2, \ldots, b_K\}$ and coup $(f, q)$ the list of features $\{c_1, c_2, \ldots, c_L\}$. Some of the features may be present on both lists, and others on only one. The two lists combined generate a third list $\{a_1, a_2, \ldots, a_M\}$. Each feature $a_m$ is active in a cognitive state $r_m$ that one or more subjects collapses to under either context $e$ or $f$. By calculating the relative frequencies of these features, we obtain an estimate of $\mu(r_m, e, p)$ and $\mu(r_m, f, q)$ The distance between $p$ and $q$ is now defined as follows:

$$d_p(q,c,f,p) = \frac{\sqrt{2}}{\sqrt{M}} \sqrt{\sum_{m=1}^{M} (\mu(r_m, c, p) - \mu(r_m, f, q))^2} \qquad (17)$$

We call $d_p$ the probability property distance, or the $p$ distance, to distinguish it from $d_\mu$, the probability distance or $\mu$ distance.

Remark that to compare this distance $d_p$ to the $\mu$ distance $d_\mu$ we introduce the renormalization factor $\sqrt{2}/\sqrt{M}$. This is to make the maximal distance, which is attained if $|\mu(r_m, e, p) - \mu(r_m, f, q)| = 1 \forall m$, equal to $\sqrt{2}$.

We can also define a property conceptual distance based on weights of properties. Given a set of features $\{a_1, a_2, \ldots, a_M\}$, for each of $p$ and $q$, $v(p, e, a_m)$ is the weight of feature $a_m$ for $p$ under context $e$, and $v(q, f, a_m)$ is the weight of feature $a_m$ for $q$ under context $f$. The distance between states $p$ and $q$ for the two concepts under contexts $e$ and $f$ respectively can be written as follows:

$$d_w(q,e,f,p) = \frac{\sqrt{2}}{\sqrt{M}} \sqrt{\sum_{m=1}^{M} \left(v(p,e,a_m) - v(q,f,a_m)\right)^2} \qquad (18)$$

We call $d_w$ the weight property distance. It is clear that this distance depends not only on $p$ and $q$, but also on the two contexts in which the weights are obtained. How the weights depend on context follows partly from the lattice $\mathcal{L}$, which describes the relational structure of the set of features, and how this structure is related to the structure of the probability field $\mu(f, q, e, p)$ which gives the probabilities of collapse under a given context.

5.5.3. *Relationship between the two distance measurements.* It would be interesting to know whether there is a relationship between the distance measured using the probability field and the distance measured using weighted properties. In pure quantum mechanics, these two distances are equal (see Appendix III, equations (35) and (39)).

This could be tested experimentally as follows. Subjects are asked to give a single feature of a given concept. We call $e$ the context that consists of making this request. Since a concept $A$ evokes slightly different cognitive states $p$ in different subjects, they do not all respond with the same feature. Thus we obtain the set of features $\{a_1, a_2,...,a_M\}$. We denote the cognitive state of a given subject corresponding to the naming of feature $a_m$ by $p_m$. The relative frequency of feature $a_m$ gives us $\mu(p_m, e, p)$. In another experiment, we consider the same concept $A$. We consider the set of features {$a_1, a_2, ..., a_M$} collected in the previous experiment. Now subjects are asked to estimate the applicability of these features to this concept. This gives us the weight values $\nu(p, e, a_m)$. Comparing the values of $\mu(p_m, e, p)$ and $\nu(p_m, e, p)$ makes it possible to find the relation between the two distances $d_p$ and $d_a$.

## 6. Application to the pet fish problem

We now present theoretical evidence of the utility of the contextual approach using the pet fish problem. Conjunctions such as this are dealt with by incorporating context dependency, as follows: (i) activation of **pet** still rarely causes activation of **guppy**, and likewise (ii) activation of **fish** still rarely causes activation of **guppy.** But now (iii) **pet fish** causes activation of the potentiality state **pet** *in the context of* **pet fish** AND **fish** in the context of **pet fish**. Since for this potentiality state, the probability of collapsing onto the state **guppy** is high, it is very likely to be activated.

### 6.1. *The probability distance*

Let us now calculate the various distance measures introduced in the previous section. We use equation (15) for the relevant states and contexts involved:

$$d_\mu(q, e, p) = \sqrt{2(1 - \sqrt{\mu(q, e, p)})} \tag{19}$$

where $\mu(q, e, p)$ is the probability that state $p$ changes to state $q$ under the influence of context $e$. Two states and three contexts are at play if we calculate the different distances $d_m$ for the pet fish situation. State $p$ is the cognitive state of a subject before any question is asked. Contexts $e$, $f$, and $g$ correspond to asking subjects to give an example of **pet, fish** and **pet fish** respectively. State q corresponds to the cognitive state consisting of the concept **guppy.**

The transition probabilities are $\mu(q, e, p)$, the probability that a subject answers 'guppy' if asked to give an example of **pet,** $\mu(q, f, p)$, the probability that the subject answers 'guppy' if asked to give an example of **fish,** and m $\mu(q, g, p)$ the probability that the subject answers 'guppy' if asked to give an example of **pet fish.** The probability distances are then:

$$d_\mu(q, e, p) = \sqrt{2(1 - \sqrt{\mu(q, e, p)})} \tag{20}$$

$$d_\mu(q, f, p) = \sqrt{2(1 - \sqrt{\mu(q, f, p)})} \tag{21}$$

$$d_\mu(q, g, p) = \sqrt{2(1 - \sqrt{\mu(q, g, p)})} \tag{22}$$

Since $\mu(q, e, p)$ and $\mu(q, f, p)$ are experimentally close to zero, while $\mu(q, g, p)$ is close to 1, we have that $d_u(q, e, p)$ and $d_u(q, f, p)$ are close to √2 (the maximal distance), and $d_u(q, g, p)$ is close to zero .

### 6.2. *The property distances*

We only calculate explicitly the weight property distance $d_w$, since this is the one calculated in representational approaches. The probability property distance $d_p$ is calculated analogously.

Four states *p, q, r, s* and four contexts *e, f, g, h* are at play. The states *p, q, r, s* are the cognitive states consisting of **guppy, pet, fish** and **pet fish** respectively. The contexts *e, f, g, h* are the experimental situations of being asked to rate the typicality of **guppy** as an instance of these four concepts respectively. For an arbitrary feature $a_m$, the weights to consider are $v(p,e,a_m), v(q,f,a_m), v(s,g,a_m),$ and $v(s,h,a_m)$. The distances are:

$$d(p,e,f,q) = \sqrt{\sum_{m=1}^{M}(v(p,e,a_m) - v(q,f,a_m))^2} \qquad (23)$$

$$d(p,e,g,r) = \sqrt{\sum_{m=1}^{M}(v(p,e,a_m) - v(r,g,a_m))^2} \qquad (24)$$

$$d(p,e,h,s) = \sqrt{\sum_{m=1}^{M}(v(p,e,a_m) - v(s,h,a_m))^2} \qquad (25)$$

Thus we have a formalism for describing concepts that is not stumped by a situation wherein an entity that is neither a good instance of *A* nor *B* is nevertheless a good instance of *A* AND *B*. Note that whereas in representational approaches, relationships between concepts arise through overlapping context-independent distributions, in the present approach, the closeness of one concept to another (expressed as the probability that its potentiality state will collapse to an actualized state of the other) is context-dependent . Thus it is possible for two states to be far apart with respect to a one context (for example $d_\mu(q,e,p)$, the distance between **guppy** and the cognitive state of the subject prior to the context of being asked to name a pet), and close to one another with respect to another context (for example $d_\mu(q,g,p)$, the distance between **guppy** and the cognitive state of the subject prior to the context of being asked to name a pet fish).

## 7. **Summary and Conclusions**

Representational theories of concepts-such as prototype, exemplar and schemata or theory-based theories-have been adequate for describing cognitive processes occurring in a focused, evaluative, analytical mode, where one analyses relationships of cause and effect. However, they have proven to be severely limited when it comes to describing cognitive processes that occur in a more intuitive, creative, associative mode, where one is sensitive to and contextually responds to not just the most typical properties of an item, but also less typical (and even hypothetical or imagined) properties. This mode evokes relationships of not causation, but correlation, such that new conjunctions of concepts emerge spontaneously. This issue of conjunctions appears to have thrown a monkey wrench into concepts research, but we see this as a mixed blessing. It brought to light two things that have been lacking in this research: the notion of 'state' and a rigorous means of coping with potentiality and context.

First a few words about the notion of 'state'. In representational approaches, a concept is represented by one or more of its states. A prototype, previously encountered exemplar, or theory description merely one state of a concept. The competition between different representational approaches seems to boil down to 'which of the states of a concept most fully captures the potentiality of the concept'? Since different experimental designs elicit different context-specific instantiations of a concept, it is not surprising that the states focused on in one theory have greater predictive power in some experiments, while the states focused on in another theory have greater predictive power in others. The true state of affairs, however, is that none of the states can represent the whole of the concept, just as none of the states of a billiard ball can represent the whole of the billiard ball. The billiard ball itself is described by the structure of the state space, which includes all possible states, given the variables of interest and how they could change. If one variable is location and another velocity, then each location-velocity pair constitutes a state in this state space. To represent the whole of an entity-whether it be a concept or a physical object-one needs to consider the set of all states, and the structure this set has.

This is the motivation for describing the essence of a concept as a potentiality state. The potentiality state can, under the influence of a context, collapse to a prototype, experienced exemplar, or an imagined or counterfactual instance. The set of all these states, denoted $\Sigma A$ for a concept $A \in \mathcal{A}$, is the state space of concept $A$. It is this state space $\Sigma A$, as a totality, together with the set of possible contexts $\mathcal{M}$, and these two sets structured within the SCOP $(\Sigma A, \mathcal{M}, \mathcal{L}, \mu, \nu)$ that represents the concept. Hence a concept is represented by an entire structure—including the possible states and their properties, and the contexts that bring about change from one state to another—rather than by one or a few specific state(s).

This brings us to the notion of context. If a theory is deficient with respect to its consideration of state and state space, it is not unlikely to be deficient with respect to the consideration of context, since contexts require states upon which to act. The contextualized approach introduced here makes use of a mathematical generalization of standard quantum mechanics, the rationale being that the problems of context and conjunction are very reminiscent to the problems of measurement and entanglement that motivated the quantum formalism. Below we summarize how these two problems manifest in the two domains of physics and cognition, and how they are handled by quantum mechanics and its mathematical generalizations.

- **The measurement problem for quantum mechanics.** To know the state of a micro-entity, one must observe or measure some property of it. However, the context of the measurement process itself changes the state of the micro-entity from superposition state to an eigenstate with respect to that measurement. Classical physics does not incorporate a means of modeling change of state under the influence of context. The best it can do is to avoid as much as possible any influence of the measurement on the physical entity under study. However, the change of state under the influence of a measurement context—the quantum collapse—is explicitly taken into account in the quantum mechanical formalism. The state prior to, and independent of, the measurement, can be retrieved as a theoretical object-the unit vector of complex Hilbert space-that reacts to all possible measurement contexts in correspondence with experimental results. Quantum mechanics made it possible to describe the real undisturbed and unaffected state of a physical entity even if most of the experiments that are needed to measure properties of this entity disturb this state profoundly (and often even destroy it).

- **The measurement problem for concepts.** According to Rips' No Peeking Principle, we

cannot be expected to incorporate into a model of a concept how the concept interacts with knowledge external to it. But *can* a concept be observed, studied, or experienced in the absence of a context, something external to it, whether that be a stimulus situation or another concept? We think not. We adopt a Peeking Obligatory approach; concepts require a peek-a measurement or context-to be elicited, actualized, or consciously experienced. The generalization of quantum mechanics that we use enables us to explicitly incorporate the context that elicits a reminding of a concept, and the change of state this induces in the concept, into the formal description of the concept itself. The concept in its undisturbed state can then be 'retrieved' as a superposition of its instantiations.

- **The entanglement problem for quantum mechanics.** Classical physics could successfully describe and predict relationships of causation. However, it could not describe the correlations and the birth of new states and new properties when micro-entities interact and form a joint entity. Quantum mechanics describes this as a state of entanglement, and use of the tensor product gives new states with new properties.

- **The entanglement problem for concepts.** Representational theories could successfully describe and predict the results of cognitive processes involving relationships of causation. However, they could not describe what happens when concepts interact to form a conjunction, which often has properties that were not present in its constituents. We treat conjunctions as concepts in the context of one another, and we investigate whether the relative SCOP might prove to be the algebraic operation that corresponds to conjunction.

Note that the measurement/peeking problem and the entanglement/conjunction problem both involve context. The measurement/peeking problem concerns a context very rxternal to, and of a different sort from, the entity under consideration: an observer or measuring apparatus in the case of physics, and a stimulus in the case of cognition. In the entanglement/conjunction problem, the context is the same sort of entity as the entity under consideration: another particle in the case of physics, or another concept in the case of cognition. The flip side of contextuality is potentiality; they are two facets of the more general problem of describing the kind of nondeterministic change of state that takes place when one has incomplete knowledge of the universe in which the entity (or entities) of interest, and the measurement apparatus, are operating.

     The formalisms of quantum mechanics inspired the development of mathematical generalizations of these formalisms such as the State Context Property system, or SCOP, with which one can describe situations of varying degrees of contextuality. In the SCOP formalism, pure classical structure (no effect of context) and pure quantum structure (completely contextual) fall out as special cases. Applying the SCOP formalism to concepts, pure analytic (no effect of context) and pure associative (completely contextual) modes fall out as special cases. In an analytic mode, cognitive states consist of pre-established concepts. In an associative mode, cognitive states are likely to be potentiality states (i.e. not collapsed) with respect to contexts. This can engender a recursive process in which the content of the cognitive state is repeatedly reflected back at the associative network until it has been completely defined in terms of some conjunction of concepts, and thus potentiality gets reduced or eliminated with respect to the context. Eventually a new stimulus context comes along for which this new state is a superposition state, and the collapse process begins again. It has been proposed that the onset of

the capacity for a more associative mode of thought is what lay behind the origin of culture approximately two million years ago (Gabora 1998, submitted), and that the capacity to shift back and forth at will from analytical to associative thought is what is responsible for the unprecedented burst of creativity in the middle/upper Paleolithic (Gabora, submitted).

We suggest that the reason conjunctions of concepts can be treated as entangled states is because of the presence of EPR-type correlations among the properties of concepts, which arise because they exist in states of potentiality, with the presence or absence of particular properties of a concept being determined *in the process of* evoking or actualizing it. If, concepts are indeed entangled, and thus for any concept, given the right context, any feature could potentially become involved in its collapse, then the notion of conceptual distance loses some meaning. What *can* be defined is not the distance between concepts, but the distance between *states* of them.[10] That said, the measure $d_\mu$ determines the distance between the cognitive state prior to context (hence a potentiality state) to the state after the influence of context (hence the collapsed state). The measure $d_p$ determines distance between two potentiality states. Note that the distance measures used in the prototype and exemplar models are actually distances between states of concepts, not between concepts themselves. This means that the distances we introduce are no less fundamental or real as measures of conceptual distance.

Preliminary theoretical evidence was obtained for the utility of the approach, using the pet fish problem. Conjunctions such as this are dealt with by incorporating context-dependency, as follows: (i) activation of **pet** still rarely causes activation of **guppy,** and likewise (ii) activation of **fish** still rarely causes activation of **guppy.** But now (iii) **pet fish** causes activation of the superposition state **pet** *in the context of* **pet fish** AND **fish** *in the context of* **pet fish.** Since for this superposition state the probability of collapsing onto the state **guppy** is high, it is very likely to be activated. Thus we have a formalism for describing concepts that is not stumped by the sort of widespread anomalies that arise with concepts, such as this situation wherein an entity that is neither a good instance of *A* nor *B* is nevertheless a good instance of the conjunction of *A* and *B*.

Despite our critique of representational approaches, the approach introduced here was obviously derived from and inspired by them. Like exemplar theory, it emphasizes the capacity of concepts to be instantiated as different exemplars. In agreement to some extent with prototype theory, experienced exemplars are 'woven together', though whereas a prototype is limited to some subset of all conceivable features, a potentiality state is not. Our way of dealing with the 'insides' of a concept is more like that of the theory or schemata approach. An instance is described as, not a set of weighted features, but a lattice that represents its relational structure. The introduction of the notion of a concept core, and the return of the notion of essence, have been useful for understanding how what is most central to a concept could remain unscathed in the face of modification to the concepts mini-theory. Our distinction between state or instantiation and potentiality state is reminiscent of the distinction between theory and core. However, the introduction of a core cannot completely rescue the theory theory until serious consideration has been given to state and context.

We end by asking: does the contextualized approach introduced here bring uscloser to an answer to the basic question 'what is a concept'? We have sketched out atheory in which concepts are not fixed representation s but entities existing in states ofpotentiality that get dynamically actualized, often in conjunction with otherconcepts, through a  collapse event that results from the interaction betweencognitive state and stimulus situation or context. But does this tell us what a conceptreally is? Just as was the case in physics a century ago, the quantum formalism, whileclearing out many troubling issues, confronts us with the limitations of science.

We cannot step outside of any particular orientation and observe directly and objectivelywhat a concept is. The best we can do is reconstruct a concepts essence from thecontextually elicited 'footprints' it casts in the cognitive states that make up a streamof thought.

**Appendices**

I.*Complex Hilbert space*

A complex Hilbert space $\mathcal{H}$is a set such that for two elements $x, y \in \mathcal{H}$of this set an operation 'sum' is defined, denoted x + y, and for any element $x \in \mathcal{H}$and any complex number $\lambda \notin C$, the multiplication of this element $x$with the complex number $\lambda$ is defined, denoted by $\lambda x$. The operation 'sum' and 'multiplication by a complex number' satisfy the normal properties that one expect these operations to satisfy (e.g. x + y = y+ x, (x+ y) + z = x+ (y+ z), $\lambda \mu x = \mu \lambda x$, etc., ... A complete list of all these properties can be found in any textbook on vector spaces). So this makes the set $\mathcal{H}$into a complex vector space, and thus we call the elements $x \in \mathcal{H}$vectors.

In addition to the two operations of 'sum' and 'multiplication by a complex number', a Hilbert space has an operation that is called the 'inproduct of vectors'. For two vectors $x, y \in \mathcal{H}$theinproduct is denoted $\langle x, y \rangle$, and it is a complex number that has the following properties. For $x, y, z \in \mathcal{H}$,and $\lambda \in \mathbb{C}$,we have:

$$\langle x, y \rangle = \langle y, x \rangle * \tag{26}$$

$$\langle x, y + \lambda z \rangle = \langle x, y \rangle + \lambda \langle, x, z \rangle \tag{27}$$

The inproduct makes it possible to define an orthogonality relation on the set of vectors. Two vectors $x, y \in \mathcal{H}$are orthogonal, denoted $x \perp y$, if and only if $\langle x, y \rangle = 0$. Suppose that we consider a subset $A \subset \mathcal{H}$,then we can introduce:

$$A^\perp = \{x | x \in \mathcal{H}, x \perp y \forall y \in A\} \tag{28}$$

which consists of all the vectors orthogonal to all vectors in A. It is easy to verify that $A^\perp$is a subspace of $\mathcal{H}$,and we call it the orthogonal subspace to A. We can also show that$A \subset (A^\perp)^\perp$,and call $(A^\perp)^\perp$,also denoted $A^{\perp\perp}$, the biorthogonal subspace of A.

There is one more property satisfied to make the complex vectorspace with an inproduct into a Hilbert space, and that is, for $A \subset \mathcal{H}$we have:

$$A^\perp + A^{\perp\perp} - \mathcal{H} \tag{29}$$

This means that for any subset $A \subset \mathcal{H}$, each vector$x \in \mathcal{H}$can always be written as the superposition:

$$x = y + z \tag{30}$$

where$y \in A^\perp$and $z \in A^{\perp\perp}$.The inproduct also introduces for two vectors $x, y \in \mathcal{H}$the measure of a distance and an angle between these two vectors as follows:

$$d(x, y) = \sqrt{\langle x - y, x - y \rangle} \tag{31}$$

$$\theta = \arccos|\langle x, y \rangle| \tag{32}$$

and for one vector $x \in \mathcal{H}$, the measure of a length of this vector:

$$\|x\| = \sqrt{\langle x, y \rangle} \tag{33}$$

This distance makes the Hilbert space a topological space (a metric space). It can be shown that for $A \subset \mathcal{H}$, we have that $A^\perp$ is a topologically closed subspace of $\mathcal{H}$, and that the biothogonal operation is a closure operation. Hence $A^{\perp\perp}$ is the closure of $A$. This completes the mathematical definition of a complex Hilbert space.

**II.** *Quantum mechanics in Hilbert space*
In quantum mechanics, the states of the physical entity under study are represented by the unit vectors of a complex Hilbert space $\mathcal{H}$. Properties are represented by closed subspaces of $\mathcal{H}$, hence subsets that are of the form $A^{\perp\perp}$ for some $A \subset \mathcal{H}$. Let us denote such closed subspaces by $M \subset \mathcal{H}$, and the collection of all closed subspaces by $\mathcal{P}(\mathcal{H})$. For a physical entity in a state $x \in \mathcal{H}$, where $x$ is a unit vector, we have that property $M$ is 'actual' if and only if $x \in M$. Suppose that we consider a physical entity in a state $x \in \mathcal{H}$ and a property $M \in \mathcal{P}(\mathcal{H})$ that is not actual, hence potential. Then, using (30), we can determine the weight of this property. Indeed there exists vectors $y, z \in \mathcal{H}$ such that:

$$x = y + z \tag{34}$$

and $y \in M$ and $z \in M^\perp$. We call the vector $y$ the projection of $x$ on $M$, and denote it $P_M(x)$, and the vector $z$ the projection of $x$ on $M$, and denote it $P_{M^\perp}(x)$. The weight $v(x, M)$ of the property $M$ for the state $x$ is then given by:

$$v(x, M) = \langle x, P_M(x) \rangle \tag{35}$$

The vectors $\frac{y}{\|y\|}$ (or $\frac{P_M(x)}{\|P_M(x)\|}$) and $\frac{z}{\|z\|}$ (or $\frac{P_{M^\perp}(x)}{\|P_{M^\perp}(x)\|}$) are also called the collapsed vectors under measurement context $\{M, M^\perp\}^M$. An arbitrary measurement context $e$ in quantum mechanics is represented by a set of closed subspaces $\{M_1, M_2, \ldots, M_n, \ldots\}\}$ (eventually infinite), such that:

$$M_i \perp M_j \, \forall i \neq j \tag{36}$$

$$\sum_i M_i = \mathcal{H} \tag{37}$$

The effect of such a measurement context $\{M_1, M_2, \ldots, M_n, \ldots\}$ is that the state $x$ that the physical entity is in when the measurement context is applied collapses to one of the states:

$$\frac{P_{M_i}(x)}{\|P_{M_i}(x)\|} \tag{38}$$

and the probability $\mu(P_{M_i}(x), e, x)$ of this collapse is given by:

$$\mu(P_{M_i}(x), e, x) = \langle x, P_{M_i}(x) \rangle \tag{39}$$

If we compare (35) and (39) we see that for a quantum mechanical entity the weight of a property $M$ for a state $x$ is equal to the probability that the state $x$ will collapse to the state $\frac{P_M(x)}{\|P_M(x)\|}$, if the measurement context $\{M, M^\perp\}$ is applied to this physical entity in this state. That is the reason that it would be interesting to compare these quantities in the case of concepts (see section 5.5.3).

### III. SCOP systems of pure quantum mechanics.

The set of states $\Sigma_Q$ of a quantum entity is the set of unit vectors of the complex Hilbert space $\mathcal{H}$. The set of contexts $\mathcal{M}_Q$ of a quantum entity is the set of measurement contexts, i.e. the set of sequences $\{M_1, M_2, \ldots, M_n, \ldots\}$ of closed subspaces of the Hilbert space $\mathcal{H}$, such that:

$$M_i \perp M_j \,\forall i \neq j \tag{40}$$

$$\sum_i M_i = \mathcal{H} \tag{41}$$

Such a sequence is also called a *spectral family*. The word spectrum refers to the set of possible outcomes of the measurement context under consideration. In quantum mechanics, a state $p \in \Sigma_Q$ changes to another state $q \in \Sigma_Q$ under the influence of a context $e \in \mathrm{M}_Q$ in the following way. If $\{M_1, M_2, \ldots, M_n, \ldots\}$ is the spectral family representing the context $e$, and $x$ the unit vector representing the state $p$, then $q$ is one of the unit vectors:

$$\frac{P_{M_i}(x)}{\|P_{M_i}(x)\|} \tag{42}$$

and the change of $x$ to $\frac{P_{M_i}(x)}{\|P_{M_i}(x)\|}$ is called the quantum collapse. The probability of this change is given by:

$$\mu(e, q, e, p) = \langle x, P_{M_i}(x) \rangle \tag{43}$$

Remark that in quantum mechanics the context $e$ is never changed. This means that:

$$\mu(f, q, e, p) = 0 \,\forall f \neq e \tag{44}$$

As a consequence, we have for the reduced probability (see (12)):

$$\mu(q, e, p) = \mu(e, q, e, p) = \langle x, P_{M_i}(x) \rangle \tag{45}$$

A property *a* of a quantum entity is represented by a closed subspace *M* of the complex Hilbert space $\mathcal{H}$. A property *a* represented by *M* always has a unique orthogonal property $a^\perp$ represented by $M^\perp$ the orthogonal closed subspace of *M*. This orthogonal property $a^\perp$ is the quantum-negation of the property *a*. The weight *v(p,a)* of a property *a* towards a state *p* is given by:

$$v(p,a) = \langle x, P_{M_i}(x)\rangle \tag{46}$$

where *M* represents *a* and *x* represents *p*. Remark that at first sight, the weight does not appear to depend on a context, as it does for a general state context property system. This is only partly true. In pure quantum mechanics, the weights only depend on context in an indirect way, namely because a property introduces a unique context, the context corresponding to the measurement of this property. This context is represented by the spectral family $\{M, M^\perp\}$.

**IV.** *SCOP systems applied to cognition*
A state context property system ($\Sigma$, $\mathcal{M}$, $\mathcal{L}$, $\mu$, $\nu$) consists of three sets $\Sigma$, $\mathcal{M}$ and $\mathcal{L}$ and two functions $\mu$ and $\nu$.

$\Sigma$ is the set of cognitive states of the subjects under investigation, while $\mathcal{M}$ is the set of contexts that influence and change these cognitive states. $\mathcal{L}$ represents properties or features of concepts. The function $\mu$ is defined from the set $\mathcal{M} \times \Sigma \times \mathcal{M} \times \Sigma$ to the interval [0, 1] of real numbers, such that:

$$\sum_{f\in\mathcal{M}, qz\in\Sigma} \mu(f,q,e,p) = 1 \tag{47}$$

and $\mu(f,q,e,p)$ is the probability that the cognitive state *p* changes to cognitive state *q* under influence of context *e* entailing a new context *f*.

We noted that properties of concepts can also be treated as concepts. Remark also that it often makes sense to treat concepts as features. For example, if we say 'a dog is an animal', it is in fact the feature 'dog' of the object in front of us that we relate to the feature 'animal' of this same physical object. This means that a relation like 'dog is animal' can be expressed within the structure $\mathcal{L}$ in our formalism.

This relation is the first structural element of the set $\mathcal{L}$, namely a partial order relation, denoted <. A property $a \in \mathcal{L}$ 'implies' a property $b \in \mathcal{L}$, and we denote $a < b$, if only if, whenever a is true then also *b* is true. This partial order relation has the following properties. For $a, b, \in \mathcal{L}$ we have:

$$a < a \tag{48}$$

$$a < b \text{ and } b < a \Longrightarrow a = b \tag{49}$$

$$a < b \text{ and } b < c \Longrightarrow a = c \tag{50}$$

For a set of properties $\{a_i\}$ there exists a conjunction property denoted $\wedge_i a_i$. This conjunction property $\wedge_i a_i$ is true if and only if all of the properties $a_i$ are true. This means that for $a_i, b \in \mathcal{L}$ we have:

$$b < \wedge_i a_i \Leftrightarrow b < a_i \forall i \tag{51}$$

The conjunction property defines mathematically an infimum for the partial order relation <. Hence we demand that each subset of $\mathcal{L}$ has an infimum in $\mathcal{L}$, which makes $\mathcal{L}$ into a *complete lattice*.

Each property $a$ also has the 'not' (negation) of this property, which we denote $a^\perp$. This is mathematically expressed by demanding that the lattice $\mathcal{L}$ be equipped withan orthocomplementation, which is a function from $\mathcal{L}$ to $\mathcal{L}$ such that for $a, b, \in \mathcal{L}$ we have:

$$(a^\perp)^\perp = a \tag{52}$$

$$a < b \Rightarrow b^\perp < a^\perp \tag{53}$$

$$a \wedge a^\perp = 0 \tag{54}$$

where 0 is the minimal property (the infimum of all the elements of $\mathcal{L}$), hence a property that is never true. The makes $\mathcal{L}$ into a complete ortho-complemented lattice.

The function $v$ is denned from the set $\Sigma \times \mathcal{M} \times \mathcal{L}$ to the interval [0, 1], and $v(p, e, a)$ is the weight of property $a$ under context $e$ for state $p$. For $a \in \mathcal{L}$ we have:

$$v(p, e, a) + v(p, e, a^\perp) = 1 \tag{55}$$